\documentclass[aps,pra,twocolumn,floatfix,superscriptaddress]{revtex4}

\usepackage[utf8]{inputenc}
\usepackage{amsmath,amssymb,amsfonts,bbm,graphicx,times,color,mathptmx,hyperref}

\newcommand{\ket}[1]{\left\vert {#1} \right\rangle} 
\newcommand{\bra}[1]{\left\langle {#1} \right\vert}

\begin{document}
 
\title{Phase estimation with squeezed single photons}

\author{Stefano Olivares}
\email{stefano.olivares@fisica.unimi.it}
\affiliation{Quantum Technology Lab, Dipartimento di Fisica "Aldo Pontremoli", Universit\`a degli Studi di
Milano, I-20133 Milano, Italy}
\affiliation{INFN Sezione di Milano, I-20133 Milano, Italy}

\author{Maria Popovic}
\affiliation{Quantum Technology Lab, Dipartimento di Fisica "Aldo Pontremoli", Universit\`a degli Studi di
Milano, I-20133 Milano, Italy}

\author{Matteo G. A. Paris}
\affiliation{Quantum Technology Lab, Dipartimento di Fisica "Aldo Pontremoli", Universit\`a degli Studi di
Milano, I-20133 Milano, Italy}
\affiliation{INFN Sezione di Milano, I-20133 Milano, Italy}

\begin{abstract}
We address the performance of an interferometric setup in which a squeezed single photon interferes at a beam splitter with a coherent state. Our analysis in based on both the quantum Fisher information and the sensitivity when a Mach-Zehnder setup is considered and the difference photocurrent is detected at the output. We compare our results with those obtained feeding the interferometer with a squeezed vacuum (with the same squeezing parameter of the squeezed single photon) and a coherent state in order to have the same total number of photons circulating in the interferometer. We find that for fixed squeezing parameter and total number of photons there is a threshold of the coherent amplitude interfering with the squeezed single photon above which the squeezed single photons outperform the performance of squeezed vacuum (showing the highest quantum Fisher information). When the difference photocurrent measurement is considered, we can always find a threshold of the squeezing parameter (given the total number of photons and the coherent amplitude) above which squeezed single photons can be exploited to reach a better sensitivity with respect to the use of squeezed vacuum states also in the presence of non unit quantum efficiency.
\end{abstract}

\maketitle

\section{Introduction}
The use of nonclassical resources, such as single photons and squeezed light,
can improve the sensitivity to a phase shift of optical interferometers also in the
presence of real setup and detectors affected by losses \cite{raf2,grav:11,DD:13,RB:13,RB:15}.
In particular, it is well known that adding squeezing at the input of an interferometer
can lead to the Heisenberg limit \cite{par95}, namely, the ultimate bound to precision allowed by
the very laws of quantum mechanics \cite{pez08}. In particular, in the last years many efforts have
been made to investigate the ultimate limits to precisions addressing different scenarios
\cite{oli:par:OptSp,lan13,lan14,CS:JOSAB,raf:rev,CS:16}.
Though squeezed states play a relevant role in practical interferometry, the peculiar
features of single-photon states allow better investigating the fundamental aspects of
the phenomenon \cite{scia:10,macro:ent}.
\par
In this paper we consider a squeezed single photon (SqSPh) and a coherent state (CS) as inputs
of interferometer and we study the behaviour of the resulting sensitivity
to detect a phase shift. Since a SqSPh can be generated starting from a squeezed
vacuum state (SqVac) by means of the photon subtraction technique \cite{wen:05,oli:par},
it is natural to compare the results to case of a SqVac and a CS as inputs. However,
it is worth noting that this is not the optimal case, which is instead achieved when squeezing is
present at both the input ports of the interferometer \cite{CS:JOSAB}. Here, we are interested
in comparing the performance of the two scenarios when the squeezed parameter and total number
of photons circulating in the interferometer are fixed. First of all we study the quantum Fisher
information for the two configurations (SqSPh+CS and SqVac+CS) given the constraints and then
we evaluate the sensitivity in the case of Mach-Zehnder interferometer where the measured quantity
is the difference between the two output photocurrents. We also consider the effect of a non unit
quantum efficiency.
\par
The paper is structured as follows. In Section~\ref{s:model} we introduce the model of an interferometer
and of the Mach-Zehnder interferometer. We also review the basic elements of the quantum estimation
theory focusing, in particular, on the Fisher and quantum Fisher information and the sensitivity of
the interferometer consider throughout the the paper. In Section~\ref{s:QFI} we show the results
concerning the quantum Fisher information whereas the sensitivity, also in the presence of non
unit quantum efficiency, is studied in Section~\ref{s:sen}. Finally, Section~\ref{s:concl} draws some
concluding remarks.

\section{The interferometer and quantum estimation theory}\label{s:model}

\begin{figure}[tb!]
\begin{center}
\includegraphics[width=0.9\columnwidth]{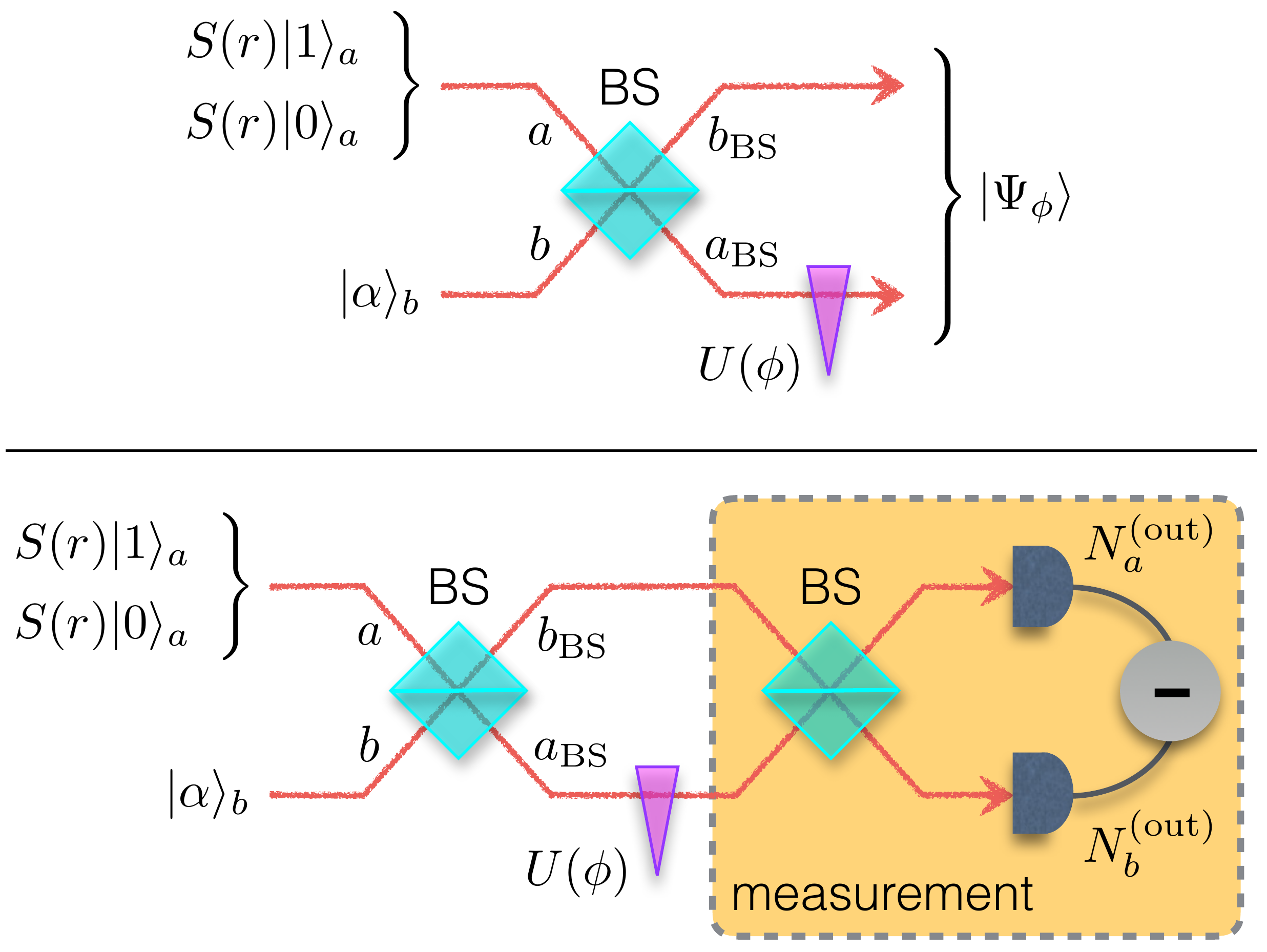}
\end{center}
\vspace{-0.6cm}
\caption{\label{f:scheme} (Top) Scheme of the interferometer: the two input states
$\ket{\psi}_a\otimes\ket{\alpha}_b$ interfere at a 50:50 beam splitter (BS) and
one of the two transmitted beams undergoes a phase shift $\phi$. (Bottom) Scheme
of the Mach-Zehnder interferometer: after the phase shift the two modes are mixed
at a BS and the difference photocurrent is recorded.}
\end{figure}
In our analysis we address two possible couples of states $\ket{\psi}_a\otimes\ket{\alpha}_b$ for the two input modes
$a$ and $b$ (with $[b,b^{\dag}]=[a,a^{\dag}]=\mathbbm{I}$, $\left[a,b\right]=0$), where the mode
$b$ is excited in a coherent state, whereas $\ket{\psi}_a$ can be either the SqSPh $S(r)\ket{1}_a$
or the SqVac $S(r)\ket{0}_a$, where $S(r) = \exp\left[\frac12 r ({a^{\dag}}^2 - a^2)\right]$
is the squeezing operator, as depicted in the top panel of Fig.~\ref{f:scheme}. The input modes interfere
at a 50:50 beam splitter (BS), let $a_{\rm BS} = (a+b)/\sqrt{2}$ and $b_{\rm BS} = (b-a)/\sqrt{2}$
be the Heisenberg evolution of the initial mode operators $a$ and $b$, after the
passage through BS. Then one of the modes, say $a_{\rm BS}$, 
undergoes a phase shift of amount $\phi$, described by the operator
 $U\left(\phi\right)=\exp\left(i\phi\, a_{\rm BS}^{\dagger}a_{\rm BS}\right)$, we want to estimate.
To this aim we first choose a suitable measurement, usually described by
a positive-operator-valued measurement $\{\Pi_x\}$, whose outcomes $x$
depend on the parameter $\phi$ and are distributed according to the conditional
probability $p(x|\phi) = \bra{\Psi_\phi} \Pi_x \ket{\Psi_\phi}$, $\ket{\Psi_\phi}$ being
the two-mode state coming from the interferometer (see the top panel of Fig.~\ref{f:scheme}).
Starting from the data, we define an
\emph{estimator}, namely, a function function providing the value of the $\phi$
and its variance $\Delta^2 \phi$.
\par
In classical estimation theory the Cram\'er-Rao
imposes a lower bound to variance (we drop for the sake of simplicity the statistical
scaling):
\[
\Delta^2 \phi \geq\frac{1}{F\left(\phi\right)}
\]
$F\left(\phi\right)$ being the Fisher information:
\[
F\left(\phi\right) = \int_{\Lambda} p(x|\phi)
\left[\partial_\phi \log p(x|\phi)\right]^2,
\]
where $\Lambda$ is the data sample space. However, the Cram\'er-Rao refers
to the actual chosen measurement. Using the tools of quantum estimation
theory \cite{par:QEQT}, we can look for the optimal measurement minimising
the uncertainty or, equivalently, maximising the Fisher information. Therefore,
we can introduce the so-called quantum Fisher information \cite{hel76,bro9x}:
\[
Q_{\rm F}\left(\phi\right)=\mbox{Tr}\left[\rho_\phi\, L_{\phi}^{2}\right],
\]
where $\rho_\phi = \ket{\Psi_\phi}\bra{\Psi_\phi}$ and $L_{\phi}$ is the symmetric
logarithmic derivative, $\partial_\phi \rho_{\phi} =
(L_{\phi}\rho_{\phi} + \rho_{\phi} L_{\phi})/2$. By definition,
$Q_{\rm F}\left(\phi\right) \ge F\left(\phi\right)$, thus we obtain the
quantum Cram\'er-Rao bound \cite{bra94,bra96}:
\[
\Delta^2\phi\geq\frac{1}{Q_{\rm F}\left(\phi\right)}.
\]
Since we are addressing a family of pure states which come to depend on the
parameter $\phi$ through a unitary operator of the form $U_{\phi}=\exp(-i\phi G)$, where
$G=a^\dag_{\rm BS}a_{\rm BS}$ is the Hermitian generator, the quantum Fisher information
can be evaluated as \cite{par:QEQT}:
\begin{equation}\label{QF:def}
Q_{\rm F}=4\left\langle \Psi_{\rm in}\right|\Delta^{2}G\left|\Psi_{\rm in}\right\rangle ,
\end{equation}
$\left|\Psi_{\rm in}\right\rangle = \left|\psi\right\rangle_a\otimes \left|\alpha\right\rangle_b$
being the quantum state entering the interferometer (see the top panel of Fig.~\ref{f:scheme}),
which is thus independent of $\phi$.
\par
Up to now we have considered the optimal scenario based on the optimal measurement.
However, in practice one should choose a particular detection scheme, according to the
current technology. In the bottom panel of Fig.~\ref{f:scheme} we depict a typical
Mach-Zehnder interferometer, where during the measurement stage
the two modes interfere at a second BS before a photodetection process,
which measures the difference photocurrent between the two output modes
$a_{\rm out}$ and $b_{\rm out}$, namely:
\begin{equation}
O(\phi) = \bra{\Psi_{\rm in}}N_a^{\rm (out)}-N_b^{\rm (out)}\ket{\Psi_{\rm in}},
\end{equation}
with $N_k^{\rm (out)} = k^{\dag}_{\rm out} k_{\rm out}$, $k= a,b$.
It is worth noting that given a small fluctuation $\delta \phi$, we can write:
\[
O(\phi+\delta\phi) \approx
O(\phi) +\partial_\phi O(\phi)\,\delta\phi,
\]
and, thus, we have the following change of the photocurrent difference:
\[
O(\phi+\delta\phi) - O(\phi) \approx \partial_\phi O(\phi)\,\delta\phi.
\]
In order to detect such a difference we should require that
$\left[ O(\phi+\delta\phi) - O(\phi) \right]^2 \gtrsim \Delta^2 O(\phi)$ or, equivalently,
$| \partial_\phi O(\phi)\, \delta\phi | \gtrsim \sqrt{\Delta^2 O(\phi)}$. Therefore, there is a
minimum value that can be detected by the apparatus, which is 
the sensitivity of the interferometer given by:
\begin{equation}\label{eq:1x}
s\left(\phi\right)=\frac{\sqrt{\Delta^2 O(\phi)}}{\left|\partial_\phi O(\phi)\right|}.
\end{equation}
It is possible to show \cite{CS:16} that the sensitivity is lower bounded by the
inverse of the Fisher information associated with the measurement, and we
have:
\begin{equation}
s\left(\phi\right)\gtrsim\frac{1}{\sqrt{F\left(\phi\right)}}\geqslant\frac{1}{\sqrt{Q_{\rm F}}}.
\end{equation}
\par
In the following we will evaluate the quantum Fisher information and the Fisher
information considering as input states a SqSPh or a SqVac and a CS and
we will compare the performance of the interferometer.

\section{Quantum Fisher information}\label{s:QFI}

In order to have the same squeezing factor and total number of photons $N_{\rm tot}\ge 1$,
we rewrite the two two-mode input states as follows (without loss of generality we can
assume the squeezing parameter $r$ and the CS amplitude $\gamma$ to be real):
\begin{subequations}
\begin{align}
\ket{\Psi_{\rm in}^{\rm (SqSPh)}} &= S(r)\ket{1}_a\otimes \ket{\sqrt{N_{\rm tot}-(\cosh 2r+\sinh^2 r)}}_b,\\[1ex]
\ket{\Psi_{\rm in}^{\rm (SqVac)}} &= S(r)\ket{0}_a\otimes \ket{\sqrt{N_{\rm tot}-\sinh^2 r}}_b,
\end{align}
\end{subequations}
or:
\begin{subequations}\label{inputs}
\begin{align}
\ket{\Psi_{\rm in}^{\rm (SqSPh)}} &= S(r)\ket{1}_a\otimes \ket{\gamma}_b,\\[1ex]
\ket{\Psi_{\rm in}^{\rm (SqVac)}} &= S(r)\ket{0}_a\otimes \ket{\sqrt{\gamma^2 + \cosh 2r}}_b,
\end{align}
\end{subequations}
where we introduced the (real) coherent amplitude $\gamma$, so that $N_{\rm tot} = \gamma^2 +\cosh 2r + \sinh^2 r$.
The second parametrisation can be more useful since, in a typical setup, one fixes the squeezing parameter $r$
and the CS amplitude $\gamma$ (note that in order to have the same $N_{\rm tot}$ the CS which interferes
with the SqVac should have a larger energy than the one interfering with the SqSPh). 
\par
\begin{figure}[h!]
\begin{center}
\includegraphics[width=0.6\columnwidth]{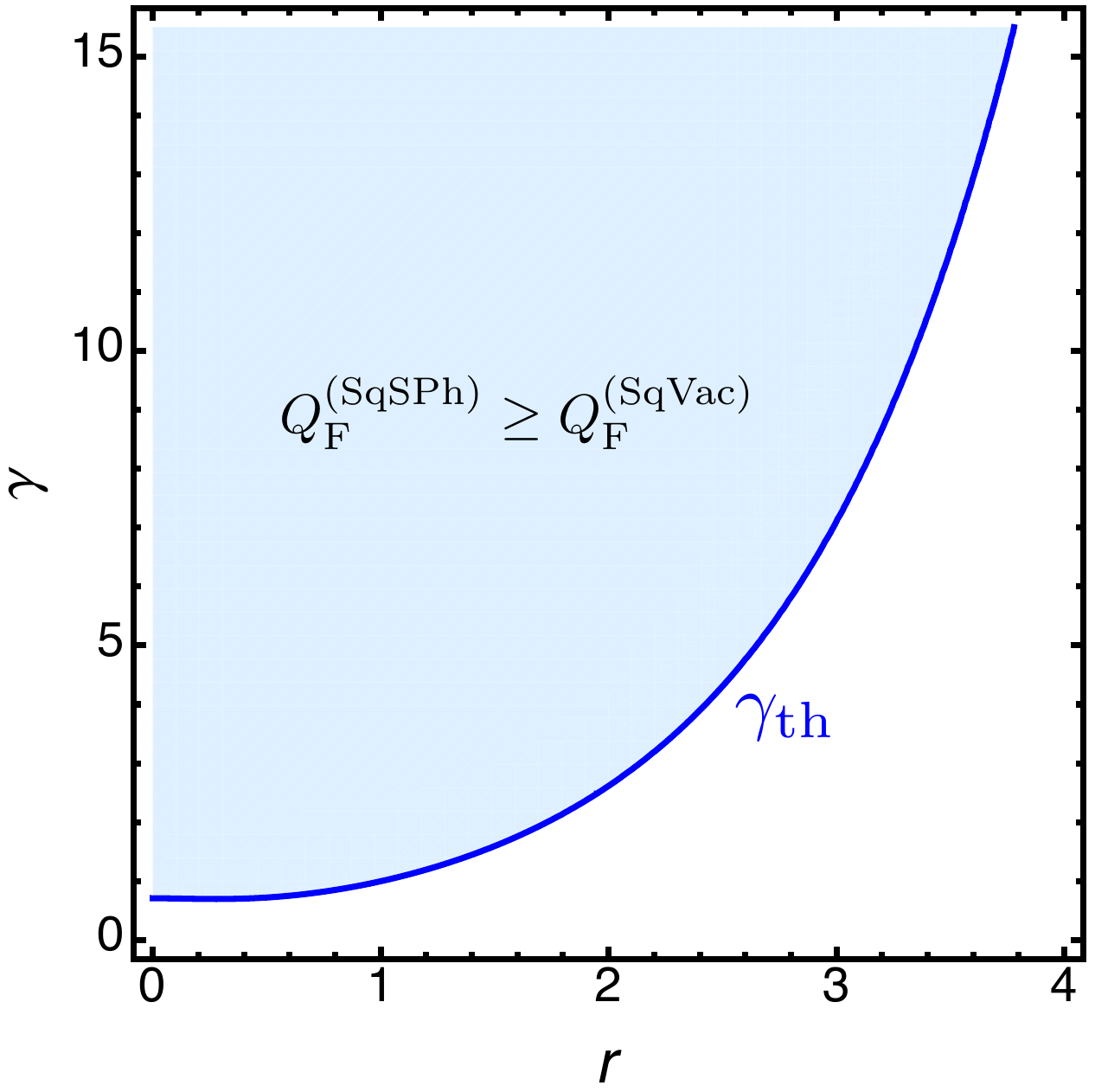}
\end{center}
\vspace{-0.6cm}
\caption{\label{f:QF:contour} Plot of the threshold $\gamma_{\rm th}(r)$: for $\gamma > \gamma_{\rm th}(r)$
we have $Q_{\rm F}^{\rm (SqSPh)}(\gamma,r) \ge Q_{\rm F}^{\rm (SqVac)}(\gamma,r)$ (shaded region).
See the text for details.}
\end{figure}
Exploiting Eq.~(\ref{QF:def}) and Eqs.~(\ref{inputs}), we can compare the quantum Fisher information
in the two cases, namely $Q_{\rm F}^{\rm (SqSPh)}(\gamma,r)$ and $Q_{\rm F}^{\rm (SqVac)}(\gamma,r)$.
Though the calculation is quite straightforward, the analytical results are cumbersome and
they are not explicitly reported here; we just observe that the quantum Fisher information is maximised
for $\phi = \pi/2$ and this will be our working point throughout the rest of the paper.
\begin{figure}[t!]
\begin{center}
\includegraphics[width=0.8\columnwidth]{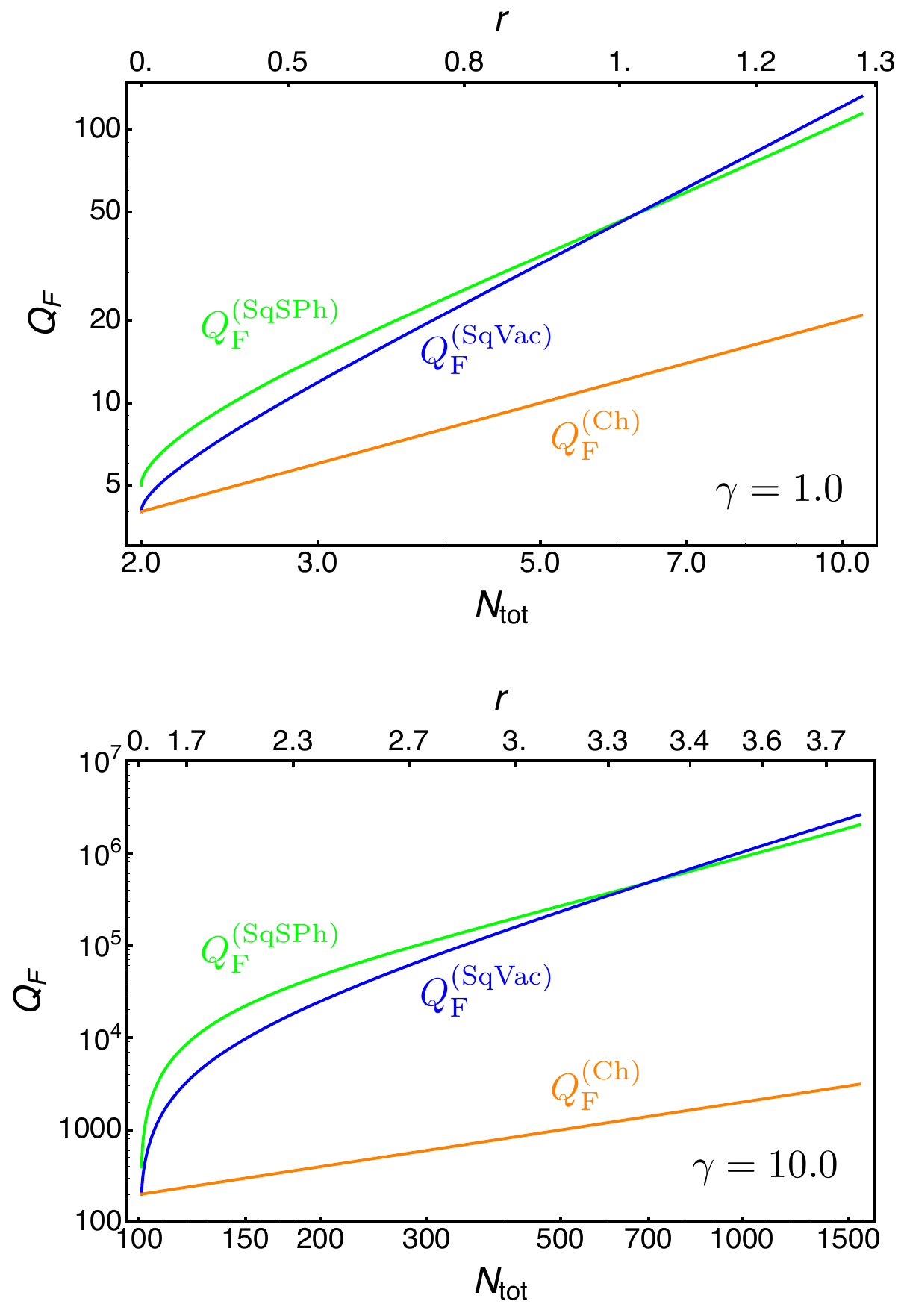}
\end{center}
\vspace{-0.5cm}
\caption{\label{f:QF:gamma} Plots of the of $Q_{\rm F}^{\rm (SqSPh)}$ and  $Q_{\rm F}^{\rm (SqVac)}$
as functions of $N_{\rm tot}$ (lower horizontal axis) or $r$ (upper horizontal axis) for two values of
the coherent amplitude: $\gamma = 1.0$ (top panel) and $\gamma = 10.0$ (bottom panel). Note
that increasing $N_{\rm tot}$ corresponds to add squeezing to the system. As expected from
Fig.~\ref{f:QF:contour}, we can identify a low energy regime where
$Q_{\rm F}^{\rm (SqSPh)} \ge Q_{\rm F}^{\rm (SqVac)}$. As $N_{\rm tot}$ gets larger the Heisenberg
scaling $\propto N_{\rm tot}^2$ is reached. For comparison, we also show the quantum Fisher
information $Q_{\rm F}^{\rm (Ch)} = 2 N_{\rm tot}$ (orange lines) referring to a single coherent state
mixed with the vacuum (in this case the upper axis is meaningless). See the text for details.}
\end{figure}
In Fig.~\ref{f:QF:contour} we plot the region of the $r \gamma$--plane
for which $Q_{\rm F}^{\rm (SqSPh)}(\gamma,r) \ge Q_{\rm F}^{\rm (SqVac)}(\gamma,r)$: given the
squeezing parameter $r$ there is a threshold value
\[
\gamma_{\rm th}(r) = \frac12 {\rm e}^{-r} \sqrt{2 + \sinh 4r},
\]
such that for $\gamma > \gamma_{\rm th}(r)$ the SqSPh outperforms SqVac.
It is worth noting that for each point in Fig.~\ref{f:QF:contour} the quantum Fisher information
$Q_{\rm F}^{\rm (SqSPh)}(\gamma,r)$ and $Q_{\rm F}^{\rm (SqVac)}(\gamma,r)$ refer to states with
the same $N_{\rm tot}$ according to the parametrisation in Eqs.~(\ref{inputs}).
In Fig.~\ref{f:QF:gamma} we plot
the two quantum Fisher information as functions of $N_{\rm tot}$ (or $r$) and fixed value of the coherent amplitude
$\gamma$. In this cases we have the following asymptotic
behaviour in the high number of photons limit $N_{\rm tot} \gg 1$ (or large squeezing parameter $r$): 
\begin{align}
Q_{\rm F}^{\rm (SqSPh)} &\approx \frac{2}{3} N_{\rm tot}^2, \\[2ex]
Q_{\rm F}^{\rm (SqVac)} &\approx \frac{10}{9} N_{\rm tot}^2,
\end{align}
respectively, that is in both the cases we find the Heisenberg scaling as one may expect
\cite{pez08,CS:JOSAB}.
It is worth noting that, at least in the presence of the optimal measurement,
the squeezing resource allows outperforming the coherent light. This is
clear form Fig.~\ref{f:QF:gamma}, where we also show the behaviour of the quantum
Fisher information $Q_{\rm F}^{\rm (SqSPh)} = 2 N_{\rm tot}$ for a coherent state mixed
with the vacuum.

\section{Sensitivity}\label{s:sen}
\par
\begin{figure}[t!]
\begin{center}
\includegraphics[width=0.6\columnwidth]{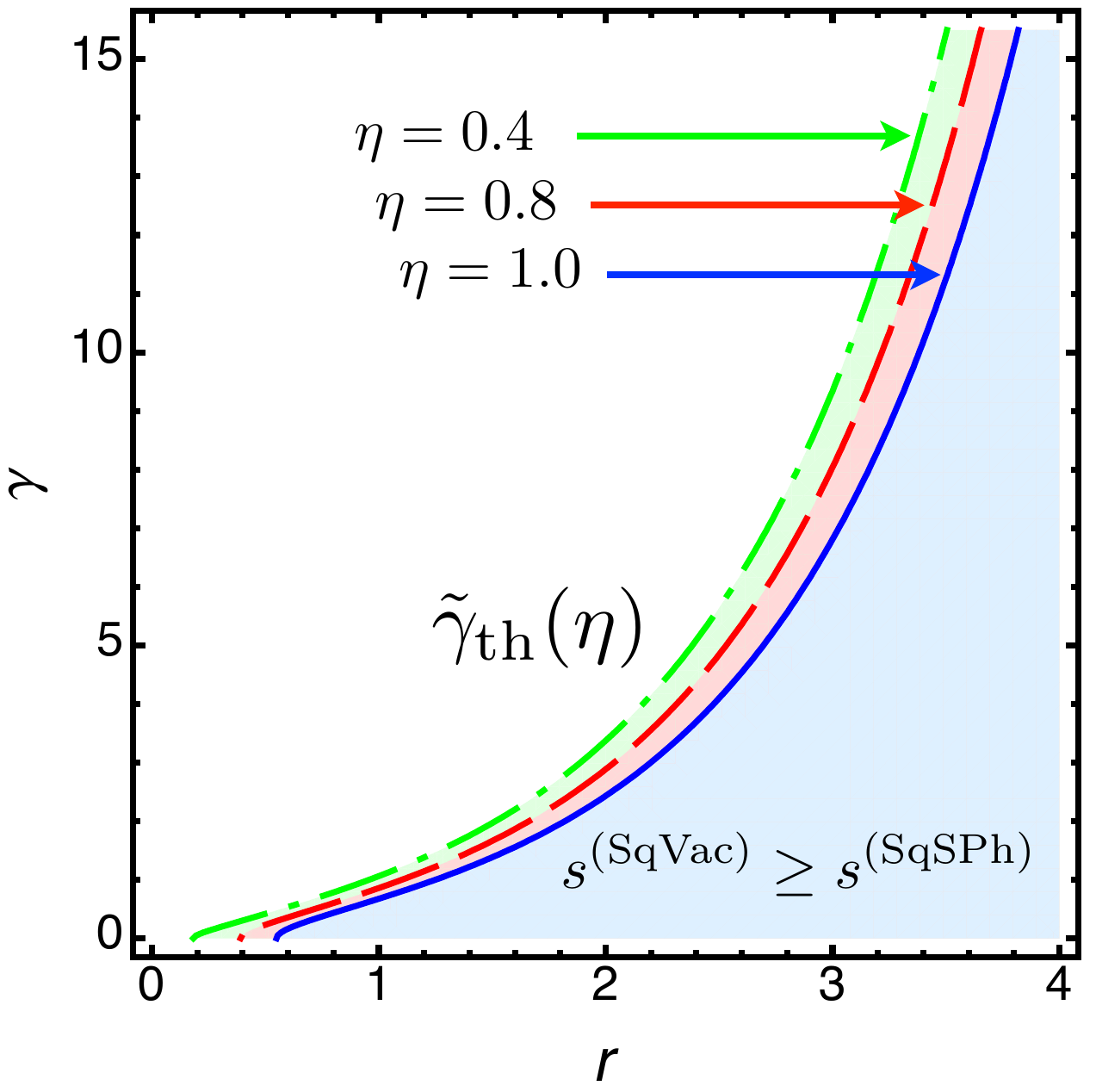}
\end{center}
\vspace{-0.6cm}
\caption{\label{f:s:contour} Plot of the threshold $\tilde{\gamma}_{\rm th}(\eta)$: for $\gamma \le \tilde{\gamma}_{\rm th}(\eta)$
we have $s^{\rm (SqVac)}(\gamma,r;\eta) \ge s^{\rm (SqSPh)}(\gamma,r;\eta)$ (shaded regions). The colors refers to
different values of the quantum efficiency. Note that the lower is the quantum efficiency, the larger is the
region in which SqSPh performs better than SqVac. See the text for details.}
\end{figure}
In this section we address the sensitivity of the Mach-Zehnder interferometer setup sketched in the
bottom panel of Fig.~\ref{f:scheme}. As in the case concerning the
quantum Fisher information, also the calculation of the sensitivity, as defined in Eq.~(\ref{eq:1x}),
can be straightforwardly obtained starting from the input states (\ref{inputs}). The analytical results
are clumsy and they are not reported explicitly.
\par
\begin{figure}[t!]
\begin{center}
\includegraphics[width=0.8\columnwidth]{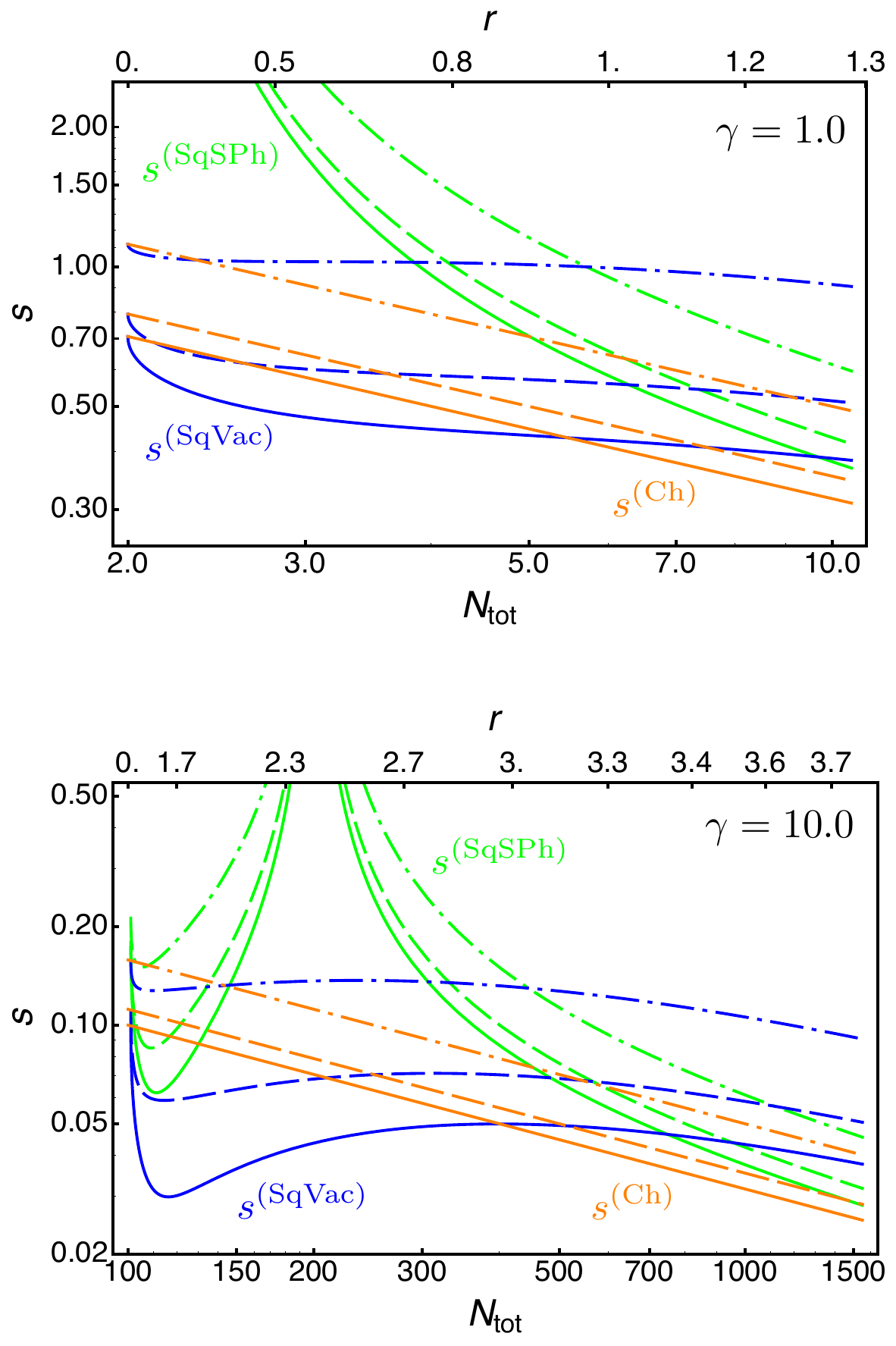}
\end{center}
\vspace{-0.5cm}
\caption{\label{f:s:gamma} Plots of the sensitivity $s^{\rm (SqSPh)}$ (green lines) and  $s^{\rm (SqVac)}$ (blue lines)
as functions of $N_{\rm tot}$ (lower horizontal axis) or $r$ (upper horizontal axis) for two values of
the coherent amplitude ($\gamma = 1.0$ (top panel) and $\gamma = 10.0$ (bottom panel)) and different
values of the quantum efficiency: $\eta = 1.0$ (solid lines), $\eta = 0.8$ (dashed lines) and $\eta = 0.4$ (dot-dashed lines).
Note that increasing $N_{\rm tot}$ corresponds to add squeezing to the system. As expected from
Fig.~\ref{f:s:contour}, we can identify a high energy regime where
$s^{\rm (SqSPh)} \le s^{\rm (SqVac)}$ (SqSPh performs better). As $N_{\rm tot}$ gets larger the shot-noise
scaling $\propto N_{\rm tot}^{-1/2}$ is reached. For comparison, we also plotted  the sensitivity
$s^{\rm (Ch)}=(\eta N_{\rm tot})^{-1/2}$ (orange lines) referring to a coherent state mixed with
the vacuum (in this case the upper axis is meaningless).
See the text for details.}
\end{figure}
In Fig.~\ref{f:s:contour} we plot the sensitivities $s^{\rm (SqSPh)}(\gamma,r;\eta)$ and
$s^{\rm (SqVac)}(\gamma,r;\eta)$, where $\eta$ is the quantum efficiency of the photodetectors
\cite{FOP:05}; the comparison is obtained for fixed total number of photons $N_{\rm tot}$ (we recall
that $\gamma$ is the amplitude of the CS interfering with the SqSPh, therefore, in general,
the two configurations have the same total energy, same squeezing parameter $r$ but different coherent
amplitude). With respect to the quantum Fisher information (see Fig.~\ref{f:QF:contour}), we can see that for fixed $r$
now we have a threshold $\tilde{\gamma}_{\rm th}$ of the coherent amplitude \emph{below} which SqSPh outperforms
SqVac. Moreover, as the quantum efficiency becomes lower, the actual value of $\tilde{\gamma}_{\rm th}$
increases: losses at the detection are more detrimental for a setup based on SqVac.
\par
In Fig.~\ref{f:s:gamma} we plot $s^{\rm (SqSPh)}$ and  $s^{\rm (SqVac)}$ for two fixed values of $\gamma$.
In the same plots we also report $s^{\rm (Ch)} = (\eta N_{\rm tot})^{-1/2}$, that is the sensitivity obtained when
a coherent state with amplitude $\sqrt{N_{\rm tot}}$ and the vacuum state are considered as inputs.
In this case we find the following scaling for the high photon number regime $N_{\rm tot}\gg 1$ (or large squeezing
parameter $r$):
\begin{align}
s^{\rm (SqSPh)} &\approx \frac{1}{\sqrt{ \eta\, N_{\rm tot}}}  \, \equiv  s^{\rm (Ch)}, \\[2ex]
s^{\rm (SqVac)} &\approx \sqrt{\frac{3(3-2 \eta)}{\eta \, N_{\rm  tot}}},
\end{align}
respectively, that is in both the case we find the shot-noise limit $\propto N_{\rm tot}^{-1/2}$:
in this limit the SqSPh performs better that SqVac.
\par
\begin{figure}[t!]
\begin{center}
\includegraphics[width=0.8\columnwidth]{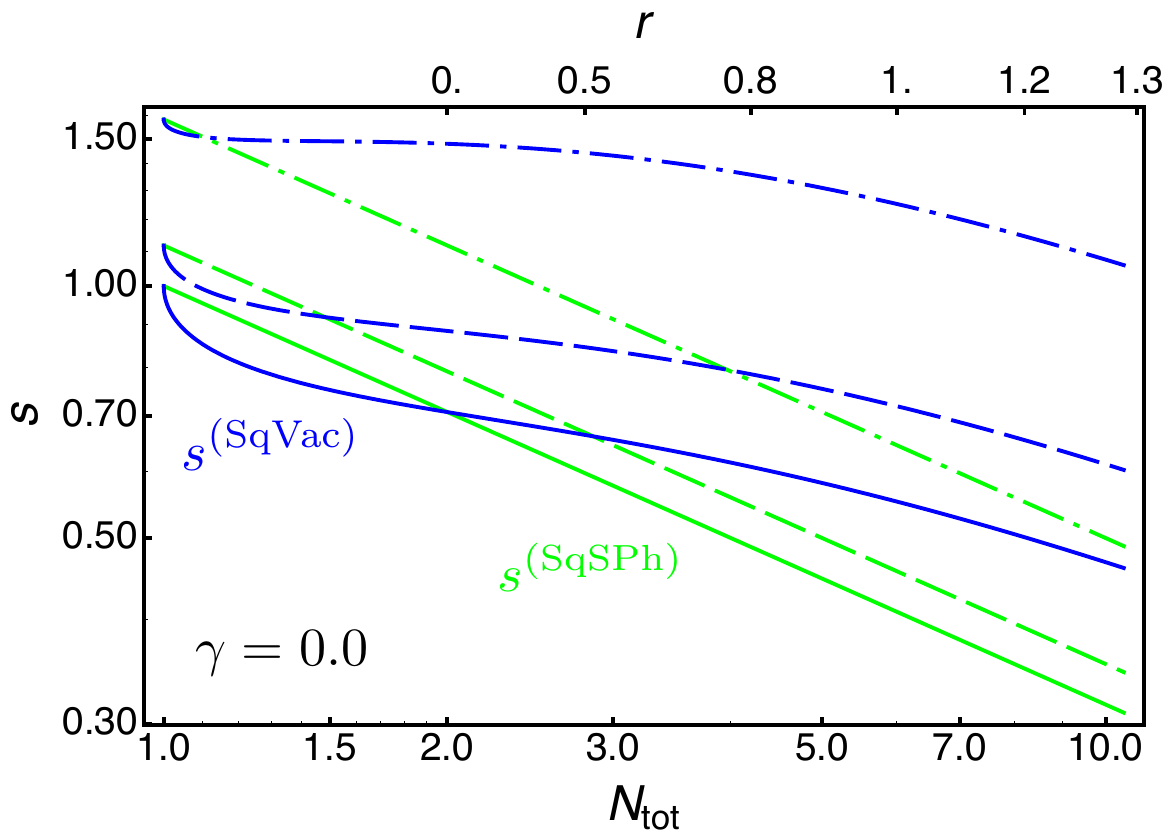}
\end{center}
\vspace{-0.5cm}
\caption{\label{f:s:gamma:vacuum} Plots of the sensitivity $s^{\rm (SqSPh)}$ (green lines) and  $s^{\rm (SqVac)}$ (blue lines)
as functions of $N_{\rm tot}$ (lower horizontal axis) or $r$ (upper horizontal axis) for $\gamma = 0.0$ and different
values of the quantum efficiency: $\eta = 1.0$ (solid lines), $\eta = 0.8$ (dashed lines) and $\eta = 0.4$ (dot-dashed lines).
Note that we have $s^{\rm (SqSPh)} = s^{\rm (Ch)} = (\eta N_{\rm tot})^{-1/2}$.}
\end{figure}
Inspecting Fig.~\ref{f:s:contour}, it is interesting to note when the SqSPh is mixed with
the vacuum ($\gamma = 0$), there is a minimum value of the squeezing parameter $r$ above
which a SqSPh allows reaching a better sensitivity than a setup exploiting SqVac
mixed with a suitable CS in order to have the same $N_{\rm tot}$ (see also Fig.~\ref{f:s:gamma:vacuum}).
However, in this last case one have $s^{\rm (SqSPh)} = s^{\rm (Ch)} = (\eta N_{\rm tot})^{-1/2}$,
namely, the squeezed single photon performs as a coherent state with the same energy.

\section{Conclusions}\label{s:concl}
In this manuscript we have investigated the performance of a SqSPh
as a probe to detect some optical phase shift. We have carried out our analysis
comparing the results from the interference of the SqSPh with a CS and with the results
obtained addressing a SqVac. In particular we focused on the case of fixed squeezing
parameter (assumed to be the same for the SqSPh and the SqVac) and fixed total
number of photons. Addressing both the quantum Fisher information and the
Mach-Zehnder interferometer (based on photodetectors), we have found the regimes
in which a SqSPh can outperform a SqVac as input. Our results show that whereas
in the optimal case, i.e., the case involving the optimal measurement associated with
the quantum Fisher information, both the inputs allow reaching the Heisenberg scaling
in the high energy (or squeezing) limit (though SqVac performs better), when the measurement
of the different photocurrent is considered the interferometer exploiting a SqSPh exhibits a
better sensitivity. Eventually, we also presented some results (see the top panel plots of
Figs.~\ref{f:s:gamma} and \ref{f:s:gamma:vacuum}) based on parameters that can be experimentally
reachable considering the small amount of the total energy (up ten photons) and the
reasonable amount of squeezing (below $12$~dB corresponding to $r \approx 1.38$.) \cite{DD:13}.

\section*{Acknowledgments}
This work was supported by EU through the project 
QuProCS  (Grant  Agreement No.~641277), and by UniMI
through H2020 Transition Grant No.~14-6-3008000-625.

\end{document}